# T PYXIDIS: DEATH BY A THOUSAND NOVAE

Joseph Patterson,[1,2] Arto Oksanen,[3] Jonathan Kemp,[4,2] Berto Monard,[5] Robert Rea,[6] Franz-Josef Hambsch,[7] Jennie McCormick,[8] Peter Nelson,[9] William Allen,[10] Thomas Krajci,[11] Simon Lowther,[12] Shawn Dvorak,[13] Jordan Borgman,[1] Thomas Richards,[14] Gordon Myers,[15] Caisey Harlingten,[16] & Greg Bolt[17]

[1] Department of Astronomy, Columbia University, 550 West 120th Street, New York, NY 10027; jop@astro.columbia.edu, jordan.borgman@gmail.com

[2] Visiting Astronomer, Cerro Tololo Inter-American Observatory, National Optical Astronomy Observatories, which is operated by the Association of Universities for Research in Astronomy, Inc., (AURA) under cooperative agreement with the National Science Foundation.

[3] CBA–Finland, Hankasalmi Observatory; Verkkoniementie 30, FI-40950 Muurame, Finland; arto.oksanen@jklsirius.fi

[4] Department of Physics, Middlebury College, Middlebury, VT 05753; jkemp@middlebury.edu

[5] CBA–Kleinkaroo, Klein Karoo Observatory, PO Box 281, Calitzdorp 6660, South Africa; bmonard@mweb.co.za

[6] CBA–Nelson, Regent Lane Observatory, 8 Regent Lane, Richmond, Nelson 7020, New Zealand; reamarsh@ihug.co.nz

[7] CBA–Mol, Andromeda Observatory, Oude Bleken 12, B-2400 Mol, Belgium; hambsch@telenet.be

[8] CBA–Pakuranga, Farm Cove Observatory, 2/24 Rapallo Place, Farm Cove, Pakuranga, Auckland 2012, New Zealand; farmcoveobs@xtra.co.nz

[9] CBA–Victoria, Ellinbank Observatory, 1105 Hazeldean Road, Ellinbank 3821, Victoria, Australia; pnelson@dcsi.net.au

[10] CBA–Blenheim, Vintage Lane Observatory, 83 Vintage Lane, RD 3, Blenheim 7273, New Zealand; whallen@xtra.co.nz

[11] CBA–New Mexico, PO Box 1351 Cloudcroft, NM 88317; tom_krajci@tularosa.net

[12] CBA–Pukekohe, Jim Lowther Observatory, 19 Cape Vista Crescent, Pukekohe 2120, New Zealand; simon@jlobservatory.com

[13] CBA–Orlando, Rolling Hills Observatory, 1643 Nightfall Drive, Clermont, FL; sdvorak@rollinghillsobs.org

[14] CBA–Melbourne, Pretty Hill Observatory, PO Box 323, Kangaroo Ground 3097, Victoria, Australia; tom@prettyhill.org

[15] CBA–San Mateo, 5 Inverness Way, Hillsborough, CA 94010, USA; gordonmyers@hotmail.com

[16] Caisey Harlingten Observatory, The Grange, Scarrow Beck Road, Erpingham, Norfolk NR11 7QX, United Kimgdom; caiseyh@yahoo.com

[17] CBA–Perth, 295 Camberwarra Drive, Craigie, Western Australia 6025, Australia; gbolt@iinet.net.au

**ABSTRACT**

We report a 20-year campaign to track the 1.8 hour photometric wave in the recurrent nova T Pyxidis, using the global telescope network of the Center for Backyard Astrophysics. During 1996–2011, that wave was highly stable in amplitude and waveform, resembling the orbital wave commonly seen in supersoft binaries. The period, however, was found to increase on a timescale $P/\dot{P} = 3 \times 10^5$ years. This suggests a mass transfer rate in quiescence of $\sim 10^{-7}$ $M_\odot$/yr, in substantial agreement with the *accretion* rate based on the star's luminosity. This is ~2000× greater than is typical for cataclysmic variables of that orbital period. During the post-eruption quiescence (2012–2016), the star continued on its merry but mysterious way — similar luminosity, similar $P/\dot{P}$ ($2.4 \times 10^5$ years).

The orbital signal became vanishingly weak (<0.003 mag) near maximum light of the 2011 eruption. By day 170 of the eruption, near $V = 11$, the orbital signal reappeared with an amplitude of 0.005 mag. It then gradually strengthened to its normal 0.08 mag amplitude, as the star declined to its "quiescent" magnitude of 15.7. During the ~1 year of invisibility and low amplitude, the orbital signal had increased in period by 0.0054(7)%. This is probably a measure of the mass ejected in the nova outburst. For a plausible choice of binary parameters, that mass is at least $3 \times 10^{-5}$ $M_\odot$, and probably more. This represents >300 years of accretion at the pre-outburst rate, but the time between outbursts was only 45 years. Thus the erupting white dwarf seems to have ejected at least 6× more mass than it accreted. If this eruption is typical, the white dwarf must be eroding, rather than growing, in mass. Unless the present series of eruptions is a short-lived episode, the binary dynamics appear to be a mutual suicide pact between the eroding white dwarf and the low-mass secondary, excited and rapidly whittled down, probably by the white dwarf's EUV radiation. This could be a major channel by which short-period cataclysmic variables are rapidly removed from the population — by evaporating the secondary.

# 1 INTRODUCTION

T Pyxidis is the Galaxy's most famous recurrent nova. Six times since 1890, the star has erupted to $V = 6$, and then subsided back to quiescence near $V = 15$. With spectroscopy and detailed light curves known for several of these eruptions, and with a fairly bright quiescent counterpart, T Pyx has become a well-studied star — sometimes considered a prototype for recurrent novae. Selvelli et al. (2008) and Schaefer et al. (2010) give recent reviews, and the 2011 eruption has propelled the star back into the journals with gusto (Shore 2013, Nelson et al. 2013, Schaefer et al. 2013, Chomiuk et al. 2014).

Since they are believed (and in a few cases known) to possess massive white dwarfs (WDs) accreting at a high rate, recurrent novae are a promising source for Type Ia supernovae. But since they also *eject* matter, their candidacy rests on the assumption that mass accretion in quiescence exceeds mass ejection in outburst. Estimates of these rates are notoriously uncertain, and that assumption has never undergone a significant test. A *dynamical* measure of the mass ejected, based on the precise orbital period change in outburst, would furnish the most precise and compelling evidence.

In the late 1980s, it was recognized that T Pyx might soon furnish that information, since an outburst could occur soon (optimists suggested 1988, based on the 1966 outburst and the estimated 22-year mean interval). However, the orbital period was not yet known; several photometric and spectroscopic studies gave discrepant periods, and all are now known to be incorrect.[18]

Schaefer et al. (1992) identified a persistent photometric wave with a period of 0.076 d, but discounted that as a possible orbital period, since it was not coherent from night to night. They interpreted it as a "superhump" — arising from precession of the accretion disk — and estimated an underlying $P_{orb}$ near 0.073 d. A 1996–7 observing campaign (Patterson et al. 1998, hereafter P98) revealed that the weak 0.076 d signal, difficult to discern over a single cycle, is actually quite coherent, maintaining a constant phase and amplitude over many thousands of cycles. With a precise ephemeris, it bore all the earmarks of a *bona fide* orbital period. Remarkably, that study of all timings during 1986–1997 revealed an enormous rate of period increase, with $P/\dot{P} = 3 \times 10^5$ years. Any remaining dissent from the P98 orbital-period interpretation fell away when Uthas et al. (2010, hereafter UKS) found radial-velocity variations precisely following the 0.07622 d period, but only when the same increasing-period photometric ephemeris was adopted (see their Figure 2).

This paper reports on our long-term photometric study of T Pyx with the globally distributed telescopes of the Center for Backyard Astrophysics (CBA). All the "quiescent" data are basically consistent with the P98 ephemeris (slightly tweaked). And, as hoped, the signal returned after the 2011 eruption — with a different period. Thus the sought-after dynamical measure of ejected mass may have been achieved.  We then revisit the P98 interpretation, educated by the 2011 eruption and the many recent studies of this amazing star.

---

[18] For the record, these periods are 0.1433 d (Barrera & Vogt 1989), 0.1100 or 0.0991 d (Schaefer 1990), and ~0.073 d (Schaefer et al. 1992, hereafter S92).

## 2  OBSERVATIONS

Nearly all our observations are time-series differential photometry with the worldwide CBA telescopes (Skillman & Patterson 1993; Patterson et al. 2013, hereafter P13).  Since our telescopes are small and the primary objective is detection and definition of periodic signals, most of the time series are obtained in unfiltered ("white") light, to achieve high time resolution with good signal-to-noise.  When we have many separate time series from different observatories, we use their overlaps to measure additive constants and thereby splice the data to obtain a longer time series on a common (instrumentally defined) magnitude scale. Somewhat more commonly, the various time series do not overlap, in which case we usually subtract the mean and trend to obtain a "zeroed" file.  By merging all the zeroed files, we then obtain a long time series with zero mean. This latter approach artificially blinds us to very low frequencies (below ~3 cycles/day[19]), but is optimum for the study of higher frequencies, which are our usual targets for analysis.

A summary log of observations is given in Table 1, where we have included the P98 data for completeness.  Each "night" consists of a time series from one observer — and hence there is occasional redundancy, which helps us with calibration between observatories.  During quiescence, at least 24 hours of observation were obtained during each season (except 2010), in order to accurately track any changes in period or waveform in the 0.07623 d orbital signal.  During and after the 2011 outburst, we increased the coverage substantially.  The total was 2002 hours of photometry on 497 nights during 1996–2016.

Light curves on single nights during quiescence are usually dominated by erratic flickering, illustrated by Figure 8 of P98.  For each cluster of data with dense spacing, we calculated the power spectrum, found a signal at the orbital period $P_{orb}$, and then folded on $P_{orb}$ to obtain the mean orbital light curve.  These waveforms were slightly variable but always contained a distinct dip of ~0.08 mag full amplitude.  Some of these waveforms will be shown below.  Presumably because of flickering, we found that at least 7–10 orbits were needed to obtain a stable waveform.

The periodic signal became much weaker in outburst, and we used much longer data streams (20–50 orbits) to search for the periodic signal.  It did not appear clearly until day 170 of the eruption, and then increased in strength as the star faded.  More details will be presented in Section 4.

## 3  PRE-OUTBURST (1996–2011)

---

[19] In this paper we routinely use cycles/day, or c/d, as the unit of frequency. "Day" is the natural unit of time in a long series of night-time observation, and cycles per day is our natural unit of frequency, since daily aliases are a great hazard in studies of periodic signals from our rotating planet.

Following the report of a 1.8 hour quasiperiod by S92, we made T Pyx a priority target for time-series photometry. In the 1996–7 campaign, we proved the existence of a strict 0.07623 d period, stable in phase and waveform over a 1-year baseline — and deduced a long-term cycle count which tied together timings of minima over the full 1986–1997 baseline (P98). Some doubt still remained about this cycle count; it relied on quite sparse timings earlier than 1996, and also required hypothesizing a rate of (orbital?) period change which was orders of magnitude greater than anything previously seen in cataclysmic variables.

Great stability is the main credential certifying an orbital origin, and we studied the light curves for stability and timing during each observing season since 1996. By 1999, it was clear that the P98 ephemeris was confirmed. Averaged over each dense cluster of photometry during each season, the 1.8 hour signal was stable in period, amplitude, and phase. The waveform was always close to that of Figure 10 in P98, and the timings of primary minimum tracked the P98 ephemeris, thus verifying the cycle count and the signal's consequent high stability and high $\dot{P}$. The 1996-2009 CBA increasing-period photometric ephemeris was the basis of the successful phase-up of radial velocities by UKS (their Figure 2).

Those timings of 1996–2011 minima, each averaged over 5–15 orbits, are reported in Table 2, and reduced to an O–C diagram in Figure 1. The upward curve indicates a steadily increasing period, and the good fit of the parabola is consistent with a constant rate of period change. The curve corresponds to the ephemeris

$$\text{Minimum} = \text{HJD } 2450124.831(1) + 0.0762263(2)\ E + 2.38(8)\times10^{-11}\ E^2. \quad (1)$$

This implies $dP/dt = 6.4\times10^{-10}$, or $P/\dot{P} = 3.3\times10^5$ years. This should be compared to P98's Figure 12, which included also the much sparser 1986–90 timings.

## 4 ERUPTION AND AFTERMATH

### 4.1 The Long-Awaited Eruption

The outburst was discovered and announced by Michael Linnolt on 2011 April 14 (JD 2455666; Linnolt 2011). We shall refer to this as "day 0" of the eruption, although subsequent archeology showed a brightening which started a few days earlier. Schaefer et al. (2013, hereafter S13) provides a fascinating and detailed blow-by-blow account. We then obtained time-series photometry on ~300 of the next ~500 nights, totalling ~1100 hours. We used the same techniques as at quiescence: segregate the data into dense clusters over ~4–10 nights, and look for periodic signals in each. Near maximum light, no periodic signals were found over the frequency range 3–1000 cycles/day. The full (peak-to-trough) amplitude upper limit for signals near the orbital frequency was 3–4 mmag. For frequencies above 100 c/d, the upper limit was ~1 mmag.

The first obvious detection of a periodic signal occurred when the star emerged from solar conjunction, around day 180 ($V = 11$). A 12-night time series yielded a clear signal at the orbital frequency, with an amplitude of 4 mmag. This signal grew steadily in amplitude as the star continued its decline from maximum light. Two dense clusters near day 70 (when $V = 9$, prior to solar conjunction) also produced *likely* detections of $\omega_{orb}$. (These were not statistically significant in the power spectra, but synchronous summation at $P_{orb}$ yielded the familiar waveform, and gave timings of minimum light consistent with the post-eruption ephemeris. Therefore these detections should be considered likely but not certain.)

### 4.2 Periodic Signals, Si; Aliasing, Si

After day 150, each segment showed a strong detection at $\omega_{orb}$, a weaker detection at $2\omega_{orb}$, and no other significant and repeatable signals. A typical power spectrum is shown in Figure 2a, and is substantially identical to the power spectra of quiescence. For several segments, peaks near $\omega_{orb} - 3.00$ or $2\omega_{orb} + 3.00$ cycles/day were surprisingly strong, and we briefly considered whether these might be detections of an independent signal (for which P98 suggested weak evidence, and which has been sometimes interpreted as evidence for magnetically channeled accretion). However, subtraction of the orbital signal always weakened, and usually removed, this weaker signal. This is the sign of an alias. We show the analysis in the lower frames of Figure 2. The spectral window (power spectrum of a time series with a pure sinusoid at $\omega_{orb}$ artificially inserted) is shown in Figure 2b; the close resemblance to Figure 2a suggests that the sidebands of Figure 2a's main peak are entirely a result of sampling. As a second test, we subtracted the best-fitting sinusoid at $\omega_{orb}$ from the time series, and calculated the power spectrum of the residuals. This is shown in Figure 2c. The power spectrum is pure noise, and the highest noise peak corresponds to a semi-amplitude of 0.0008 mag.

The ±3.00 cycles/day alias may be a subtle hazard of southern stars, since the southern planet is mostly water, with three major centers of astronomical research — Chile, South Africa, and Australia/New Zealand — spaced by ~120° in longitude. (*Long* nightly time series are never fooled by this distant alias, but the necessarily-short runs near solar conjunction can be.) In the historical record, this signal near 10.1 c/d has been transient and only reported in quite short segments of data, so we strongly suspect that it is just an alias. Red noise (flickering) and vast oceans are a formidable enemy in the study of periods in CVs.

It's possible to be slightly more quantitative about this. In theory, daily aliases are spaced by 1.00268 c/d, so the 3 and 4 c/d aliases in Figure 2 should occur at 10.1094(4) and 9.1067(4) c/d, assuming all observations are centered on local meridian transit. The actual peaks occur at 10.1124(14) and 9.1074(14) c/d. So the frequencies agree with the alias hypothesis within ~0.001 c/d, and the **phases** must also agree (since subtraction of the $\omega_{orb}$ signal annihilated the other signals).

Believe it or not, the power spectrum of Figure 2a really was "typical" (in its limitations, too). Because our program attempted to follow the periodic signal over years, we made many observations at times when T Pyx was not optimally placed in the night sky — forcing us to splice many relatively short runs, and thus running a bigger risk of aliases. The corresponding result for one data segment obtained in February, when T Pyx transits near local midnight, is shown in Figure 3. Here all the potential aliases are invisible, because the individual time series are long enough to bridge the longitude gaps in our network (the Atlantic, Pacific, and Indian Oceans).

### 4.3 Magnetism, Probably No

We discuss this point at length because of the many previous papers which allege or cite evidence of *magnetism* in T Pyx. No such evidence, in the form of a photometric signal at a non-orbital frequency, exists in our data. Some have also cited evidence from spectroscopy — namely the strong He II emission, sometimes thought to be associated with radial accretion. But He II emission is mainly indicative of high temperature, not the accretion geometry; and He II in T Pyx appears to be of normal strength for very luminous CVs [see Figure 5 of Patterson & Raymond (1985), where T Pyx is the point at upper right]. Photoionization by the disk's boundary-layer radiation seems capable of powering He II emission of that strength (~5 Å equivalent width). The He II emission could also arise in a wind.

Of course, there could be magnetism lurking somewhere in the star — probably 20–30% of all CVs show some magnetic effect (usually collimated accretion flow). But we know of no evidentiary basis for magnetism. Published references to magnetism in T Pyx are sufficiently numerous to be in the "proof by successive publication" zone.[20]

### 4.4 Light Curves, Timings of Minima, and Period Changes

For each segment we folded the time series (~30 orbits near $V$ = 12–13, and ~20 orbits near $V$ = 14–15.7) on $P_{orb}$, and some of the mean light curves are shown in Figure 4. The same general description applies: a roughly sinusoidal waveform, with a (relatively) large dip defining minimum light, and a small variable dip near maximum light. The waveforms are very similar to that of "pre-eruption quiescence" (Figure 10 of P98). For each segment we measured the average time of minimum light, along with the full amplitude of variation. The results are shown in Table 2. The best-fit period just after eruption is 0.0762337(1) d, an increase over the period just prior to eruption by 0.0054(7)%. So large a period change is very, very surprising: 7× larger than the $\Delta P$ predicted by Livio (1991) — and of the opposite sign!

We also looked for period change after the eruption (2011.8–2016.1). Figure 5 is an O–C diagram of the 42 timings with respect to a constant test period (0.076234 d). The upward curvature indicates a period increase, and the fitted parabola is equivalent to the elements

Minimum = HJD 2,456,234.7753(4) + 0.07623361(6) $E$ + 2.9(4)×10$^{-11}$ $E^2$.   (2)

The corresponding timescale $P/\dot{P}$ for period increase is then 2.4(4)×10$^5$ years — similar to the pre-outburst estimate in Eq. (1).

---

[20] But metaphorically, as everyone who studies T Pyx knows, the star is *hyper*-magnetic.

Unfortunately, O–C diagrams are no longer standard equipment in the astronomer's toolbox. So we show these effects more transparently in Figure 6, which tracks period versus time during 1986–2016. Each period is a 2-year running average; for example, the 2003 period is based on timings during 2002–4. Figure 6 shows the period increases before and after eruption, plus a very rapid increase which is roughly centered on the eruption (day 0 corresponding to year 2011.3).

The first two points in Figure 6 are derived from the very sparse early timings collected in Table 6 of P98. These are much less reliable, because they are mostly based on single-night light curves which "looked good". Nevertheless, the cycle count established here is identical to the P98 cycle count, and the derived ephemerides are consistent. So these early points are likely correct, although skeptical readers should feel free to ignore them.

## 5  ABSOLUTE PHASING OF THE SIGNALS

Despite the large $\Delta P$ in eruption, there is no difficulty in measuring the absolute phases across eruption. Figure 7 shows an O–C diagram of the timings for several years before and after eruption, and the straight lines indicate linear fits to the timings before and after eruption. The two lines appear to meet at day 120±90. If the $\Delta P$ occurred very rapidly, that event could have occurred at day 0, or as many as 250 days after eruption. A gradual change is also possible, of course — and is more physically plausible.

Perhaps the most interesting aspect of Figure 7 is not the exact time of the $\Delta P$ event, which is unknowable, but rather this: *the absolute phasing of minimum-light in the orbital cycle appears to be preserved across eruption* — at $V$ = 15.5, $V$ = 11, and possibly even $V$ = 8. As demonstrated in Figure 4, the shape of the light curve is also roughly preserved. Some aspect of binary structure is responsible for the orbital signal, and it seems to be basically independent of luminosity state. In §10.1 we will interpret this as a *reflection effect* in the binary, probably augmented by a small partial eclipse.

## 6  REDDENING, DISTANCE, LUMINOSITY... AND ACCRETION RATE?

In P98 we adopted a distance of 3.5±1.0 kpc and a reddening $E(B–V)$ = 0.35. This was emphatically rejected by Gilmozzi & Selvelli (2008), who obtained $E(B–V)$ = 0.25±0.02 from the standard technique of removing the λ2200 bump from the UV flux distribution. But as stressed by Fitzpatrick (1999), the scatter in the empirical relationship used to infer reddening from the λ2200 bump, even when the fluxes are very accurately known, is at least 20%. For a variable star like T Pyx it must be worse, because the two UV spectra (IUE SWP + LWR) were obtained at different times. Godon et al. (2014) revisited this subject, and their Figure 1 shows the UV spectrum combining the IUE, HST, and GALEX spectra. Figure 2 of Godon et al. shows the result of applying various reddening corrections to that combined spectrum, and the authors settle on $E(B–V)$ = 0.35 as the best choice. No error is quoted, but the figure suggests something like (–0.07,+0.10) might be realistic. The diffuse interstellar bands suggest $E(B–V)$ = 0.44±0.17 (Shore et al. 2011). Thus we consider $E(B–V)$ = 0.35 a plausible value. That estimate corresponds to a column density $N_H = 1.9 \times 10^{21}$ cm$^{-2}$, according to the correlation of Predehl & Schmitt (1995). This is also consistent with the column densities inferred from X-rays in outburst ($2.0 \times 10^{21}$ cm$^{-2}$, Chomiuk et al. 2014; $1.6 \times 10^{21}$ cm$^{-2}$, Tofflemire et al. 2013) and HI radio observations on that line of sight ($2.1 \times 10^{21}$ cm$^{-2}$, Dickey & Lockman 1990).

The mean magnitudes at "quiescence" are $V = 15.4$, $B–V = 0.07$, $U–B = –0.97$ (Landolt 1970, 1977). Assuming a normal interstellar extinction curve, the de-reddened magnitudes are then $V = 14.3$, $B–V = –0.28$, $U–B = –1.22$. These colors signify a very hot source. They're similar to the colors of a mid-O star, with $T \sim 40000$ K and a bolometric correction of $\sim 3.5$ mag (Flower 1996), and roughly the same for a model DA white dwarf (Koester, Schulz, & Weidemann 1979). Sokoloski et al. (2013) measure a distance of $4.8 \pm 0.5$ kpc from the light echoes seen in the HST images (reflecting off ejecta in the nebula). Assuming spherical symmetry and correcting for extinction with the galaxy-averaged $A_V = 3.1\ E(B–V)$, the quiescent T Pyx then has $M_V = +0.9$, $M_{bol} = –2.6$, or $L = 800\ L_\odot = 3 \times 10^{36}$ erg/s.[21]

Neglecting any contribution from a boundary layer, disk accretion onto a WD of mass near 1 $M_\odot$ yields

$$L = 3 \times 10^{35}\ m_1^{1.8}\ (\dot{M})_{18}\ \text{erg/s}, \qquad (3)$$

where $(\dot{M})_{18}$ is the accretion rate in units of $10^{18}$ g/s, and $m_1 = M_1/1\ M_\odot$ (with $m_1^{1.8}$ incorporating the WD mass-radius relation near 1 $M_\odot$). Thus we estimate

$$\dot{M} = 1 \times 10^{19}\ m_1^{-1.8}\ \text{g/s} = 1.5 \times 10^{-7}\ m_1^{-1.8}\ M_\odot/\text{yr}. \qquad (4)$$

# 7 INTERPRETATION

## 7.1 Quiescence

In quiescence, T Pyx's secondary transfers matter to the white dwarf — at a very high rate, to account for the high quiescent luminosity and the frequent nova eruptions. If total mass and angular momentum are conserved in this process, then $\dot{M}$ is related to $\dot{P}$ via

$$\dot{M} = qM_1(\dot{P}/P)/3(1-q), \qquad (5)$$

---

[21] To sharpen our analysis of the energetics, we should correct for the star's binary inclination. The star has long been regarded as nearly face-on, because the emission lines are relatively narrow and nearly stationary (UKS). But HST imaging and radial velocities of the shell ejected in 2011 is more compatible with a *high* binary inclination, and it may be possible to re-interpret the emission lines as arising in an accretion-disk wind, rather than in a rotating disk (Sokoloski et al. 2016). Also, the depth of the binary eclipse in soft X-rays (Tofflemire et al. 2013) is hard to understand with a very low inclination. We consider this to be now an open question. Inclinations above $\sim 70°$ are probably ruled out by the lack of deep eclipses and smallness of the orbital modulation... and inclinations much below $20°$ have difficulty producing much orbital modulation at all. So we'll take the coward's way out and apply no correction for inclination. (In effect, this is equivalent to adopting $i = 50–60°$).

where $M_1$ is the white-dwarf mass and $q = M_2/M_1$. For our measured $\dot{P} = 6\times10^{-10}$ (during 1996–2011, when the long baseline confers good accuracy) and the binary parameters formally deduced by UKS ($M_1 = 0.7\ M_\odot$, $q = 0.2$), this implies $\dot{M} = 1.8\times10^{-7}\ M_\odot$/yr. But the line doubling and the photometric modulations (X-ray and optical) are very surprising if the binary inclination is as low as the UKS value (10±2°). Assuming a disk-wind reinterpretation of the velocities, it is possible, though by no means certain, that the *motion* of the emission lines remains a good tracer of the true dynamical motions, although the emission-line *widths* have a completely different origin. In that case we can still use the UKS result of $v_1 \sin i = 18$ km/s to infer masses, with a dependence on the unknown inclination.

This constraint is shown in Figure 8. Of course, binary inclinations much higher than the UKS value drive $q$ much lower; in the vicinity of $i = 50$–$60°$, the solutions are near $M_1$=1.1 $M_\odot$, $M_2$=0.06 $M_\odot$. Eq. (5) then yields $\dot{M} = 6\times10^{-8}\ M_\odot$/yr.

So for a broad range of inclinations, the accretion rate inferred from the luminosity is similar to the mass transfer rate implied by the steady increase in $P_{orb}$. Both rates are near $10^{-7}$ $M_\odot$/yr, if $M_1$ is near 1 $M_\odot$. Does such a binary actually make recurrent-nova outbursts? Yes, apparently it does. With these parameters, the models of Yaron et al. (2005, their Table 3) erupt every ~80 years, with the timescale depending sharply on both $M_1$ and $\dot{M}$. Thus our physical parameters in quiescence appear to satisfy[22] all known constraints.

## 7.2 Eruption and Aftermath

During eruption, mass loss should increase $P_{orb}$, and angular-momentum loss should decrease it. It's an open question which will dominate. But our observations (Figures 6 and 7) show $\Delta P/P = +5.4\times10^{-5}$, indicating that mass loss wins. For the minimum plausible prescription for angular-momentum loss (radial ejection from the white dwarf), this implies a mass loss

$$\Delta M = 3.0\times10^{-5}\ m_1\ (1+q)\ M_\odot. \qquad (6)$$

For $m_1 \approx 1$, this represents about 300 years of accretion, yet only 45 years elapsed since the 1966 outburst. So the *prima facie* evidence suggests that the nova ejected at least 6× more matter than the WD had accreted.

One can nibble around the edges of this conclusion by revising some numbers ($m_1$, $q$, $i$, bolometric correction). It's also possible that some of the ejected matter had never been on the WD. But the assumption most susceptible to error is that the nova ejecta carry off very little angular momentum (just the specific angular momentum of the white dwarf). It's easy to imagine ways in which more angular momentum is carried away: from the secondary, from rotation, from frictional losses. But the observed $\Delta P$ is large, positive, and undeniable; so each of these would only *raise* $\Delta M$, strengthening the conclusion that the WD erodes (or at least fails to increase its mass significantly; this would be the case if much of the ejected matter never resided on the WD). We note that radio observations (from the free-free emission) also suggest a large Δ$M$, probably near $10^{-4}$ $M_\odot$ (Nelson et al. 2014). Thus it now seems unlikely that the white dwarf in T Pyx — once considered a fine ancestor for a Type Ia supernova — will ever increase its mass at all, much less reach 1.4 $M_\odot$.

---

[22] Which is not to say that we understand them! So high a mass-transfer rate from so puny a donor star is unprecedented and mysterious.

Caleo & Shore (2015) suggest an alternative hypothesis: that a change in the *eccentricity* of the binary might significantly affect the change in $P_{orb}$ — and therefore that Δ$P$ cannot be used to directly infer Δ$M$. But any change in eccentricity would presumably make only a transient effect on $P_{orb}$. As the eccentricity relaxed back to zero, $P_{orb}$ should relax back to the value appropriate for $e = 0$. Figure 6 suggests that no such relaxation is occurring. It appears that $P_{orb}$ resumes tracking the normal[23] $\dot{P}$ of quiescence, as if the eruption never happened. This probably limits the importance of eccentricity change.

## 8 T PYX AMONG THE CVs

In the ranks of CVs, T Pyx holds many titles: most luminous, hottest, highest excitation, fastest orbital-period change, most frequently erupting, etc. We have shown, or at least advocated with enthusiasm, that all of these (except perhaps "most famous") can be ascribed to just one property: highest accretion rate.

For stars powered by accretion, time-averaged $M_V$ is a good proxy for $\dot{M}$, and Figure 9 shows the empirical data on $M_V$ versus $P_{orb}$ for disk-accreting CVs of short period (<0.1 days) and "known distance".[24] Dots and triangles (which are upper limits) show garden-variety dwarf novae, and the superposed bold curve shows the prediction of the standard theory of CV evolution, in which mass-transfer is driven by angular momentum loss by gravitational radiation (GR). With a few small but systematic departures, the stars track the theory curve, resembling a boomerang, pretty well. The lighter curve, labelled GR+, shows the corresponding prediction for the slightly-enhanced angular-momentum loss rate considered by Knigge et al. (2011, $\dot{J}$ = 2.5 ($\dot{J}$)$_{GR}$ ) to improve the fit to the stellar *radii*. Squares denote a small subclass of dwarf novae known as "ER UMa stars". The N symbols indicate 20th-century novae, roughly 50 years after eruption and often assumed to be in their version of "quiescence". Two stars are shown by name: T Pyx, and BK Lyn, which is a definite ER UMa star and very likely a 2000-year-old classical nova (P13).

---

[23] Although the time baseline for this measurement is still rather short. Timings through the year 2018 will greatly improve the accuracy of this test.
[24] Readers will have a variety of opinions concerning what accuracy is required to deserve the adjective "known". More specifically, this is an expanded version of Figure 5 and Table 2 of P11, where the distance constraints are discussed — in general, and also for the individual stars. While some are high-quality distances (e.g. from trigonometric parallax or fitting of stellar-atmosphere models), most are based on standard-candle methods and only good to ~40%.

In the theory peddled by P13, that boomerang-shaped curve[25] is the main story of CV evolution, but each star experiences classical-nova eruptions, which vault the star into the upper regions, where it stays for thousands of years as *something* keeps the accretion rate high. It settles back to near-quiescence after ~50,000 years (see Figure 7 of P13 for a conjecture on the rate of decline). But some stars never get the opportunity to rest after their nova ordeals, because new classical-nova eruptions can interrupt the decline. They may happen with ever-increasing frequency, because eruption frequency scales at least as fast as $\dot{M}$ in the TNR models (see Yaron et al. 2005). Eventually the star can turn into a T Pyx, and then soon die as the secondary is evaporated after ~1000 more eruptions (0.1 $M_\odot$/10$^{-4}$ $M_\odot$).

This timescale for dropping to the CV main-sequence is discussed by P13, especially in their Figure 11. The key is to recognize that novae seem to fade logarithmically with time, roughly like $dm/d(\log t) \approx 1$. (Not $dm/dt$ = constant, which is often assumed and leads to much shorter estimates for "the end of the eruption".) This is compatible with previous Herculean studies of nova decline rate (Vogt 1990; and especially Duerbeck 1992, who had it all right). But those studies could not reach a strong conclusion because they were hampered by the short baseline available to them (~100 years). This changed with the recognition of:

(a) BK Lyn as a likely 2000-year-old nova;

(b) very faint pre-eruption magnitudes and limits (Collazzi et al. 2009, Schaefer & Collazzi 2010); and

(c) the essential difference between the short-$P_{orb}$ novae and their long-$P_{orb}$ cousins, which have strong machines ("magnetic braking") for generating luminosity unrelated to the nova event (P13).

Another constraint on timescale comes from consideration of space densities. In our census there are ~5 old novae with an average distance of ~2 Kpc, 7 ER UMas with an average distance of 400 pc, and 120 normal dwarf novae with an average distance of 250 pc. For a Galactic distribution with a vertical scale height ~300 pc, this corresponds to space densities roughly in the ratio 1:100:10000 (where "1" corresponds to 7×10$^{-10}$ pc$^{-3}$). In our interpretation, this ratio represents the time spent in these various stages. Old novae last at least 150 years (no recent short-period nova has ever become a certified dwarf nova), but probably less than 2000 years (BK Lyn is transitioning now). So we speculate that the durations of these states are roughly 10$^3$:10$^5$:10$^7$ years. That also agrees at the long end, since it requires ordinary short-period dwarf novae, accreting at 4×10$^{-11}$ $M_\odot$/yr, to erupt after accumulating 4×10$^{-4}$ $M_\odot$ — an estimate not far from that of the TNR models (10$^{-4}$ $M_\odot$, Table 2 of Yaron et al. 2005).

Could the hypothesized slowness of that decline reflect merely the cooling of the WD after outburst? Probably not. After a few years, the stars are dominated specifically by accretion light, as evidenced by all the usual signatures of accretion disks: flickering, broad eclipses, doubled emission lines, power-law flux distribution, positive and negative superhumps, etc. The feeble secondaries in these stars appear to be really transferring matter at unnaturally high rates.

---

[25] The empirical version, defined by the dots and triangles. That could perhaps be described as arising from GR plus an additional driver (angular-momentum loss, or something tracking or mimicking it) which increases with $P_{orb}$.

## 9  THE ORIGIN OF THE MASS TRANSFER

What could maintain these high rates?  A plausible mechanism is a wind from the secondary, driven by the intense EUV and supersoft X-ray radiation from the WD.  This has been previously discussed by van Teeseling & King (1998) for supersoft binaries, and specifically for T Pyx by Knigge et al. (2000).  EUV and soft X-rays will be absorbed high in the secondary's atmosphere, and therefore may have no significant direct effect on the star's structure, but all that energy can be very effective in driving a wind.  Some of the wind escapes uneventfully, and some is captured by the WD.  If the WD mass and accretion rate are high enough, nuclear burning occurs and we see a WD shining with $L_{bol} = 10^{36}$–$10^{37}$ erg/s, $T =$ 500,000 K (a "supersoft binary").  If the WD mass and/or accretion rate are somewhat lower, then accretion probably dominates the energy budget and we see a source of lower luminosity and temperature  — until years later, when the accumulated fuel burns explosively.

This is different from the popular model for supersoft sources, in which thermal-timescale mass transfer occurs from a 1–2 $M_\odot$ secondary (van den Heuvel et al. 1992).  As pointed out by Oliveira & Steiner (2007, hereafter OS), the mass ratio $q = M_2/M_1$ can in principle be tested by observing the rate of orbital period change.  For conservative mass transfer, $\dot{P} < 0$ implies $q > 1$, and $\dot{P} > 0$ implies $q < 1$ [see Eq. (5)].  Evolution changes both $q$ and $P$ rapidly, and there is also a Roche-lobe-filling constraint.  Figure 1 of OS shows the expected variation of $P/\dot{P}$ in the two basic models, and demonstrates that the sign of $P/\dot{P}$ discriminates between them.  In particular, OS argues that the  observed period increase in CAL 87 supports the theory of wind-driven mass transfer, even for that famous star — the original charter member of the supersoft binary club.  This also agrees with the masses obtained from radial velocities by Hutchings et al. (1998).

For most stars classified as supersoft binaries, the $\dot{P}$ values are unknown. Almost all are in other galaxies, and hence not visited sufficiently often by telescopes to measure $\dot{P}$. Fully-credentialed Milky Way supersofts are far less numerous, because of soft X-ray absorption in the Galactic plane. But in an important paper, Steiner & Diaz (1998) pointed out that the Milky Way contains a class of stars — which they termed "V Sge stars", a name which has stuck — which resemble the supersofts pretty thoroughly, except for the defining property of intense soft X-ray flux.  P98 also advocated the inclusion of such stars, especially the two described in that paper (V Sge and T Pyx).  All the proposed V Sge members are broadly similar to the supersofts in luminosity, accretion rate, spectrum and excitation, orbital period, orbital light curves... and they have measured $\dot{P}$s which show that they, like the supersofts, are in short-lived states. It would be quite advantageous if we can study our specimens at 11th magnitude!

Like CAL 87 and most of the V Sge stars (except V Sge itself), T Pyx has a large positive $\dot{P}$, as expected for radiation-driven winds from a low-mass secondary.

## 10 ORBITAL PHASE: WHAT DOES IT MEAN?

### 10.1 Radial-velocity and Light Variations

The UKS spectroscopic study found that the emission-line source reached superior conjunction (red-to-blue crossing) at the time of minimum light. That would be natural if minimum light arises from an *eclipse* of the accretion disk by the secondary. Indeed, a small eclipse may be present, but the orbital wave varies smoothly around the orbit, so the main effect is probably different.

The same phase relation between light and velocities is produced by a reflection effect, with the secondary heated by the intense emission from the WD and its vicinity. But if that radiation is isotropic and direct, then at most ~1% can reach the small secondary, and possibly none at all from the disk, which radiates perpendicular to the orbital plane. The *disk rim and hot spot* (where the mass-transfer stream strikes the disk) are more promising. At a high accretion rate, the disk is likely to be large (minimizing the phase offset between the spot and the secondary) and possessing a relatively high rim. Even with no help from the central object, a classical disk should be concave, with a vertical height scaling as $r^{9/8}$. This gives a favorable geometry for heating at the rim, reprocessing radiation from the central object. Add a slight bulge on the rim, presumably from stream impact, and incident radiation from disk center can puff it up further and create the substantial asymmetry on the disk that is needed to fit this hypothesis to the light curves.

This is the essence of the model which has been used with success to fit the optical light curve of the supersoft binary CAL 87 (Schandl et al. 1997, Meyer-Hoffmeister et al, 1997; see also Armitage & Livio 1998, Spruit et al. 1998). As remarked in P98 (see Figures 3, 6, and 10 of that paper), the CAL 87 optical light curve bears close resemblance to that of T Pyx, and is a "dead ringer" for that of V Sge. This appears to be substantially true for the radial velocities as well. The periodic dips in CAL 87 occur at the same time the emission-line source reaches superior conjunction (Hutchings et al. 1998; see their Table 3 and Figure 9). It's no surprise in CAL 87, because that star shows optical and X-ray eclipses, with an orbital inclination estimated as ~82° (Ribeiro et al. 2014). And in V Sge, the He II emission line reaches superior conjunction at eclipse phase 0.93±0.05 (Diaz 1999). V Sge's eclipse depths vary greatly but predictably, with minimum depth when the star is in one of its bright states. The orbital inclination is estimated as ~70° (Smak et al. 2001), permitting a true eclipse, so the observed eclipse ought to coincide (and does) with the time of conjunction.

When the orbital dip does not have the depth and overall signature of a true eclipse arising from high orbital inclination, it's harder to interpret. But we note that the light and radial-velocity variations of the supersoft binary SMC 13 = 1E 0035.4–7230, which at $P_{orb}$ = 4.1 hours is T Pyx's closest known cousin among the acknowledged supersofts, also follow this pattern — with red-to-blue crossing at photometric phase 0.00±0.03. (Crampton et al. 1997). The orbital wave in this star is approximately sinusoidal, with a full amplitude of 0.24 mag. Radial-velocities appear to track the WD in all four stars, with the photometric wave plausibly attributed to a reflection effect, and the difference in light curves (amplitude and symmetry) plausibly attributed to the accident of orbital inclination.

## 10.2 The Orbital Wave in X-rays

The soft X-rays come from a compact source, so any reasonable person would expect their orbital light curves to depend strongly on $i$: deep, short, and sharp eclipses for an edge-on binary... and very flat for binaries of lower $i$. But observations contradict this. Tofflemire et al. (2013) present the light curve for the supersoft source in T Pyx, which is a smooth sinusoid with minimum phased with the optical minimum. Ribeiro et al. (2013, see also Schmidtke et al. 1993) present the light curve of CAL 87, which also shows a smooth variation, plus at most a *partial* eclipse near optical minimum. These light curves imply that the X-ray source is quite large — an X-ray "corona", which presumably scatters X-rays coming from the WD.

## 10.3 And the Secondary...

With all this talk of a raised disk rim and only ~1% of the WD's luminosity able to reach the secondary even on a fully transparent line of sight, how does the secondary manage to be heated, as we have alleged in §9? Heating of the secondary is the linchpin of the whole machine. We don't fully, or maybe even partially, understand this. But the X-ray corona provides a promising channel. To satisfy the CAL 87 X-ray light curve, it should be comparable to the entire disk in size, and can scatter X-rays towards the donor star with no great intervening opacity.

Also, the two Δ$P$ events in T Pyx — the impulsive one in eruption, and the steady one in quiescence — imply the possibility of asynchronous rotation. Tides synchronize a binary quickly... but not as quickly as the observed Δ$P$ events may *un*synchronize this particular binary. There is a lot of energy in an M star rotating with $P$ = 1.8 hours. Tides will rapidly couple the outside of the star to the orbit, and less rapidly the inside. The resultant shear could conceivably add heat to the star — throughout, not just in its upper atmosphere.

The constraint of very low donor-star mass ($M_2$<0.06 $M_\odot$), plus the observation of steady $P_{orb}$ increase, suggest that T Pyx is a "period bouncer" — that much-discussed final phase of CV evolution. But all previous discussions (e.g. Kolb & Baraffe 1999, P11) have assumed that it occurs at *low* luminosity, totally unlike T Pyx. Still, the essence of period-bounce is that an evolutionary process is inflicted on the secondary faster than its thermal timescale. With a high luminosity, a puny secondary, and the likelihood of high heating, that appears to be very probable for T Pyx. This appears to be a separate channel of cataclysmic-variable demise, which we are only now learning because this phase is so very rapid.

## 11 QUO VADIS, T PYX?

We have now tracked the $P_{orb}$ evolution through 30 years — just about the average interval between eruptions. The observations include an eruption, and the six known eruptions are pretty close counterparts, at least in their light curves. So with a little nip from Ockham's Razor, but without proof of course, it seems reasonable to consider the possibility that this evolution will continue: with $P_{orb}$ ever increasing, each nova event carrying off ~$10^{-4}$ $M_\odot$, and progressively whittling down the secondary to smaller mass. The future would then hold ~1000 (= $10^{-1}$ $M_\odot$/$10^{-4}$ $M_\odot$) more eruptions, and then the secondary evaporates.

We have always wondered why T Pyx is unique. This scenario offers a candidate explanation: because it is dying — annihilating its secondary in a paroxysm of repeated nova events, and lasting only ~20000 more years at the current rate. Some of the population statistics of cataclysmic variables (total space densities, ratio of long-period to short-period CVs) would make more sense if there were a way to kill off short-period CVs, thereby preventing them from swamping the local census. This is "the problem of the dead novae", which has been with us for a long time (Patterson 1984, §VIIc and VIIIe; Patterson 1998, §6.3 and §6.4). It could also solve the problem of the missing period bouncers (the puzzling rarity of stars on the lower — largely invisible — part of the boomerang in Figure 9). T Pyx may offer us an embarrassingly gaudy but practical way to solve these problems.

## 12 SUMMARY

1. We provide a full report of the CBA campaign during 1996–2016, following our previous accounts (P98, P14, and the ephemeris in UKS). We acquired ~2000 hours of time-series photometry before, during, and after outburst.

2. Except for ~100 days after the rapid rise to outburst, the star always shows a distinct 0.07623 d photometric wave, with a full amplitude smoothly varying from 0.004 mag (at $V$ = 11) to 0.08 mag (at $V$ = 15.7). The waveform is nearly sinusoidal, but small and variable dips near the phase of maximum light leads us to adopt *minimum* light as the best fiducial mark of phase. We refer to the signal as "orbital", for several reasons:

    (a) No other period is found in the best-quality data sets — those with densely clustered time series and of adequate length. Binary stars may or may not have other periods, but they definitely need to have an orbital period.

    (b) The period is always present, with the same phase and amplitude (at quiescence).

    (c) The P98 ephemeris successfully predicts the minima for many years ahead.

    (d) The exact ephemeris agrees with the velocities from the UKS spectroscopic study.

(e) The photometric and spectroscopic phases agree, for a natural (ha!) choice of interpretation, where minimum light coincides with superior conjunction of the emission-line source.

3. During quiescence (1996–2011, and probably 1986–2011), the period increased smoothly on a timescale $P/\dot{P} = 3\times10^5$ years. We interpret this as due to mass transfer at a rate near $10^{-7}$ $M_\odot$/yr. This also agrees with the accretion rate required to power the luminosity.

4. Somewhere near day 120±100 days after the initial rise to outburst, the orbital period changed to 0.0762336 d — an increase of 0.0054(5) % over the value before outburst. This change, shown in Figures 6 and 7, took place over an interval of less than 1.5 years, and was probably caused by the ejection of at least $3\times10^{-5}$ $M_\odot$. This exceeds the mass transfer during the previous 45 years, suggesting that the WD mass is probably eroding. It's hard to grow WD mass in a system which has nova outbursts!

5. During 2012-16, the orbital period resumed its increase, at a rate similar to that seen during the 25 years prior to eruption. This tends to support the idea that the dominant origins of the $\Delta P$ effects are pretty straightforward: mass transfer in quiescence, mass loss in outburst.

6. The soft X-ray light curve (Tofflemire et al. 2013) shows a large orbital dip in phase with the dip in optical light. This is plausible if the binary inclination is somewhat high (>50°), and this is also suggested by the HST imaging and radial velocities of the ejected shell. A re-analysis of the UKS radial velocities at quiescence suggests that $M_2$ does not exceed 0.06 $M_\odot$ (Figure 8).

7. Figure 9 shows the $M_V$–$P_{orb}$ relation for short-period CVs which are nonmagnetic and powered by accretion. As is well known, the novae are around $M_V = 4$ and the dwarf novae are mostly around <$M_V$> = 9–10 — with the recently discovered ER UMa class around <$M_V$> = 7. We propose that every star's location above the "CV main sequence" (the boomerang) reflects mainly *time since the last nova eruption*. The total time to return to the boomerang is probably 10,000–100,000 years. Enhanced mass transfer for so long an interval produces a few strangely but briefly bright stars (old novae and ER UMas), but also can be significant in speeding up every star's long-term evolution. And if $\dot{M}$ is high enough, it can ignite new nova eruptions before the star can rest from its last one. We interpret T Pyx as an extreme example of this (hypothetical) process.

8. The linchpin of this scenario is the secondary star, which must transfer matter at unnaturally high rates. This probably occurs through a radiation-driven wind, as is thought to operate for several (most?) of the supersoft binaries.

9. The orbital light curve is unusual among accretion-powered CVs, but very similar to that of V Sge stars and several supersoft binaries. In all four stars we consider, the velocities show that the emission-line source ("disk") is at superior conjunction at the time of minimum light.

Both a reflection effect and a geometrical eclipse are consistent with this phase, and both are likely significant.

10. The estimated space density of active CVs is ~$10^{-5}$ pc$^{-3}$, and most of these are short-$P_{orb}$ stars (Patterson 1998). With a characteristic scale height of 300 pc, there should be about 300,000 CVs out to the distance of T Pyx. Any of these stars with the properties of T Pyx would be impossible to hide; some would actually be naked-eye stars (briefly). But we only know of one. We interpret this to mean that it is in a very unusual phase of its life, maybe the last phase, in which it evaporates its donor star after ~1000 more eruptions. This fate could await other short-period novae, once their secondaries become sufficiently light and less able to cope with the damage from nova events. This could be a major channel by which classical novae — and therefore all CVs — are removed from the population.

11. In our effort to understand this mysterious star, we've placed a lot of emphasis on light curves and periods. This seems to have been productive... but admittedly, "when you have a hammer, everything looks like a nail" (Maslow 1966). We now wait for others, wielding different hammers, to move the story beyond the level of comprehension we've staggered into.

The 20-year path to this paper's emergence has been made possible by a medley of grants from the National Science Foundation (most recently AST12–11129), NASA, and the Mount Cuba Astronomical Foundation. Discussions with Christian Knigge, Koji Mukai, Jeno Sokoloski, and Brad Schaefer have sharpened our ideas about novae, and occasionally stymied our speculations with uncomfortable facts. We apologize for the irregular way (rumors, private communications) the results have been leaking out — a poor model for scientific work. We'll do better next time!

# TABLE 1

## Summary Log of Observations

| Year | Nights/hours | Observers |
|---|---|---|
| 1995–1996 | 14/85 | Patterson |
| 1996–1997 | 7/29 | Patterson |
| 1997–1998 | 8/33 | Kemp |
| 1998–1999 | 8/29 | Kemp |
| 1999–2000 | 6/26 | Kemp |
| 2000–2001 | 5/24 | McCormick |
| 2001–2002 | 4/24 | Kemp, McCormick |
| 2002–2003 | 19/71 | Rea, Kemp, Monard, Allen, Richards |
| 2003–2004 | 10/39 | Rea, Allen, Monard |
| 2004–2005 | 7/24 | Rea, Monard, McCormick |
| 2005–2006 | 23/90 | Allen, Rea, Monard, Christie, McCormick |
| 2006–2007 | 30/99 | McCormick, Allen, Monard, Bolt |
| 2007–2008 | 18/74 | Rea, Bolt |
| 2008–2009 | 15/62 | Rea, Monard, McCormick, Bolt |
| 2010–2011 | 7/24 | Rea, Myers (before eruption) |
| 2011 | 102/400 | Oksanen, Harlingten. Monard, Lowther, Dvorak, Bolt, Krajci, Hambsch (eruption) |
| 2011–2012 | 54/221 | Oksanen, Harlingten, Monard, Hambsch, Myers |
| 2012–2013 | 68/243 | Oksanen, Monard, Nelson, Hambsch, McCormick |
| 2013–2014 | 15/61 | Monard, Nelson, Oksanen, Myers |
| 2014–2015 | 33/154 | Myers, Oksanen, Monard, Dvorak, Hambsch |
| 2015–2016 | 49/128 | Nelson, Monard, Myers, Hambsch, Dvorak |

| Observer | Telescope | Location/Observatory |
|---|---|---|
| Kemp | 1.0 m/0.9 m | La Serena, Chile (CTIO) |
| Oksanen | 0.5 m | San Pedro de Atacama, Chile (Caisey Harlingten Observatory) |
| Allen | 0.4 m | Nelson, New Zealand (Vintage Lane Observatory) |
| Rea | 0.35 m | Nelson, New Zealand (Regent Land Observatory) |
| McCormick | 0.35 m | Auckland, New Zealand (Farm Cove Observatory) |
| Monard | 0.35 m | Calitzdorp, South Africa (Klein-Karoo Observatory) |
| Hambsch | 0.4 m | San Pedro de Atacama, Chile (ROAD Observatory) |
| Nelson | 0.3 m | Melbourne, Australia (Ellinbank Observatory) |
| Christie | 0.4 m | Auckland, New Zealand (Auckland Observatory) |
| Bolt | 0.35 m | Perth, Australia |
| Dvorak | 0.3 m | Orlando, Florida, USA (Rolling Hills Observatory) |
| Patterson | 1.0 m/0.7 m | La Serena, Chile (CTIO) / Sutherland, South Africa (SAAO) |
| Myers | 0.42 m | Siding Spring, Australia |
| Richards | 0.3 m | Melbourne, Australia (Pretty Hill Observatory) |
| Lowther | 0.25 m | Pukekohe, New Zealand (Jim Lowther Observatory) |
| Krajci | 0.35 m | Cloudcroft, New Mexico, USA |

# TABLE 2

**Timings of Orbital Minima**

## A. Pre-outburst*

| | | Minimum | | | |
|---|---|---|---|---|---|
| | | (HJD 2,450,000+) | | | |
| 124.830 | 164.893 | 189.548 | 212.491 | 460.609 | 548.497 |
| 820.931 | 870.554 | 1171.804 | 1651.652 | 1930.872 | 2323.903 |
| 2624.9273 | 2658.9994 | 2718.9171 | 3055.3867 | 3396.9675 | 3738.0087 |
| 3751.1261 | 3770.9415 | 4113.2874 | 4144.9212 | 4152.3152 | 4178.9918 |
| 4467.0638 | 4481.9269 | 4852.9376 | 4913.9227 | 4939.3021 | 5560.9545 |
| 5588.9313 | 5598.9173 | | | | |

*Consistent with V=15.6, and amplitude 0.08 mag.

## B. Outburst and Aftermath

| Minimum (HJD 2,450,000+) | Amplitude (mag) | V | Minimum (HJD 2,450,000+) | Amplitude (mag) | V |
|---|---|---|---|---|---|
| 5835.8433 | 0.005 | 11.35 | 5879.8317 | 0.005 | 11.76 |
| 5890.7328 | 0.006 | 11.96 | 5900.7192 | 0.008 | 12.11 |
| 5909.6416 | 0.014 | 12.28 | 5924.6566 | 0.011 | 12.41 |
| 5934.6436 | 0.012 | 12.57 | 6001.6534 | 0.025 | 13.40 |
| 6034.5868 | 0.019 | 13.66 | 6063.5504 | 0.028 | 13.86 |
| 6102.5077 | 0.037 | 14.06 | 6234.7755 | 0.054 | 14.82 |
| 6255.7385 | 0.073 | 14.86 | 6265.8025 | 0.058 | 14.86 |
| 6274.7963 | 0.071 | 14.90 | 6283.5640 | 0.078 | 15.05 |
| 6288.4432 | | 15.07 | 6295.3809 | | 15.10 |
| 6298.4277 | 0.069 | 15.08 | 6302.9992 | | 15.11 |
| 6325.7200 | 0.070 | 15.10 | 6332.9637 | | 15.12 |
| 6341.5781 | 0.072 | 15.12 | 6362.1625 | | 15.18 |
| 6460.1963 | | 15.31 | 6594.8262 | | 15.44 |
| 6630.8065 | | 15.46 | 6645.3680 | 0.073 | 15.52 |
| 6653.4510 | | 15.54 | 6707.8808 | | 15.58 |
| 6789.9121 | 0.077 | 15.57 | 6979.8113 | 0.075 | 15.62 |
| 7005.4227 | 0.080 | 15.70 | 7026.0085 | | 15.71 |
| 7143.2512 | | 15.70 | 7147.2185 | | 15.71 |
| 7150.8799 | | 15.68 | 7154.8431 | | 15.68 |
| 7365.0971 | 0.080 | 15.81 | 7370.0501 | 0.081 | 15.85 |
| 7380.0412 | 0.096 | 15.84 | 7394.0682 | 0.090 | 15.85 |
| 7400.0104 | 0.073 | 15.83 | 7413.1240 | 0.072 | 15.74 |
| 7417.3196 | 0.074 | 15.73 | 7426.6939 | 0.076 | 15.79 |

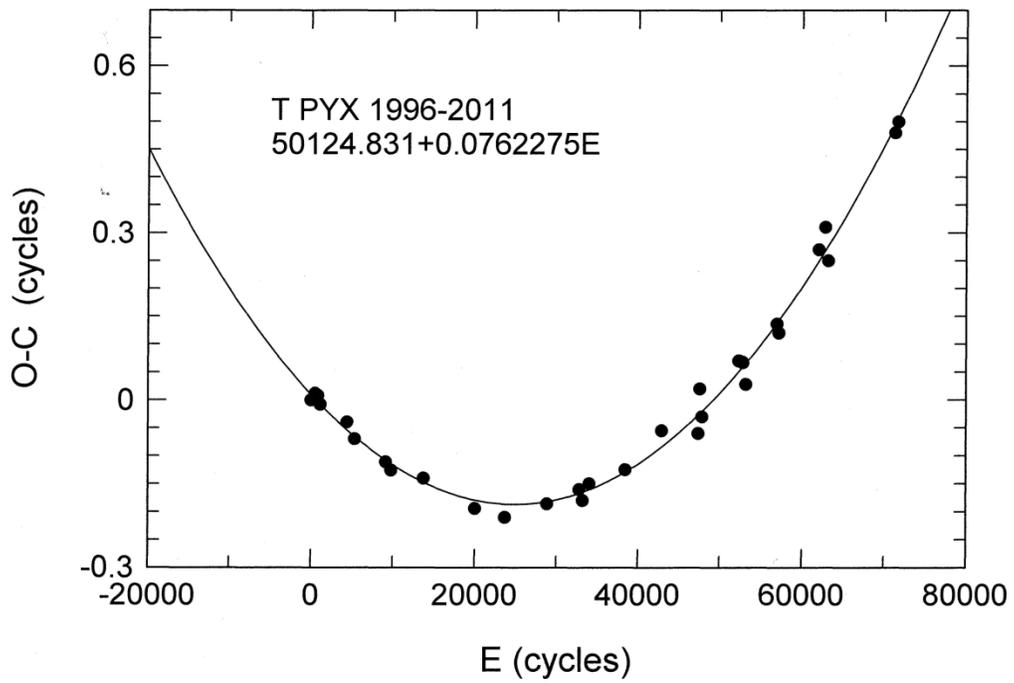

Figure 1. O–C diagram of the timings of primary minima during 1996–2011. The fit to a parabola indicates acceptable representation with a constant rate of period change ($P/\dot{P} = 3\times10^5$ years).

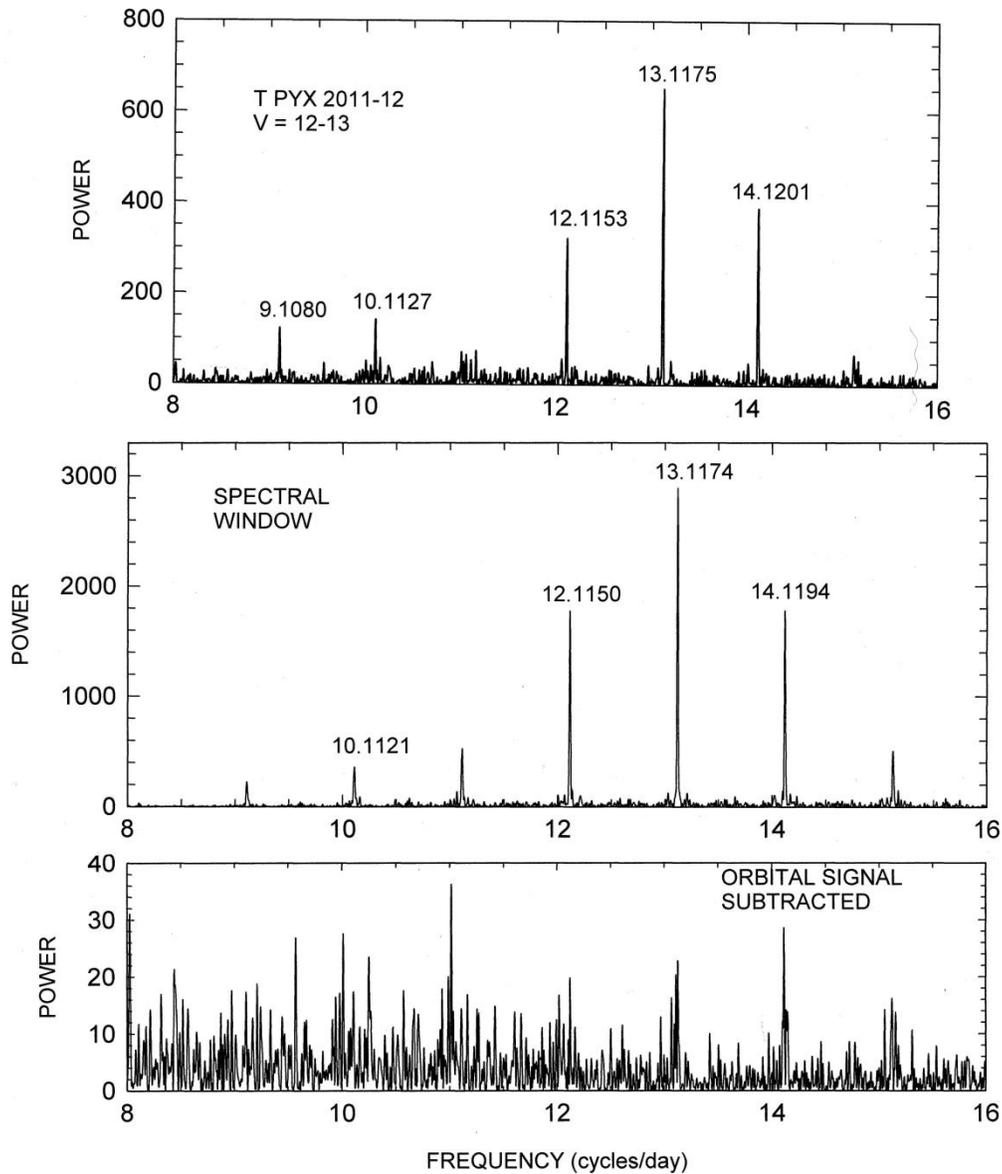

Figure 2. *Upper frame:* power spectrum of a 30-day segment of light curve soon after eruption. Signals are flagged with their frequency in cycles/day, The main signal at 13.1175 c/d appears strongly, and several likely aliases. *Middle frame:* power spectrum of an artificial time series, containing only the main signal and sampled exactly like the actual light curve. The close resemblance to the upper frame shows that the smaller peaks are entirely a result of aliasing. *Lower frame:* the power spectrum of the original time series after the main signal (period, amplitude, and phase) is subtracted, showing that no other signal is present, to a semi-amplitude upper limit of 0.6 mmag.

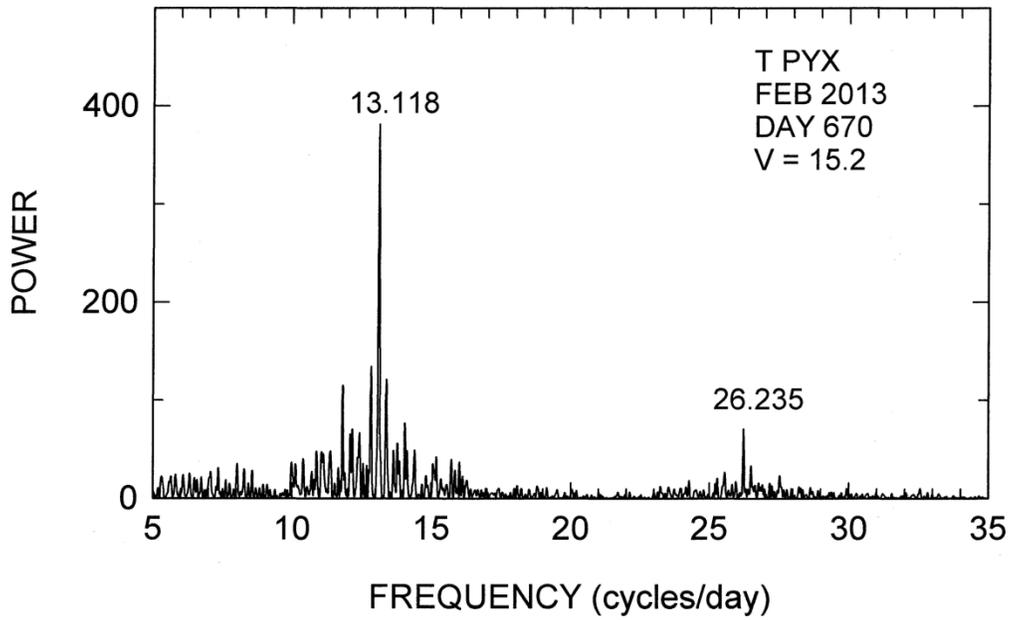

Figure 3. Power spectrum of a dense (20-day) segment of light curve, when the gaps between observing stations were minimal. The only significant signals are $\omega_{orb}$ and its harmonics; upper limits at other frequencies are typically ~4 mmag.

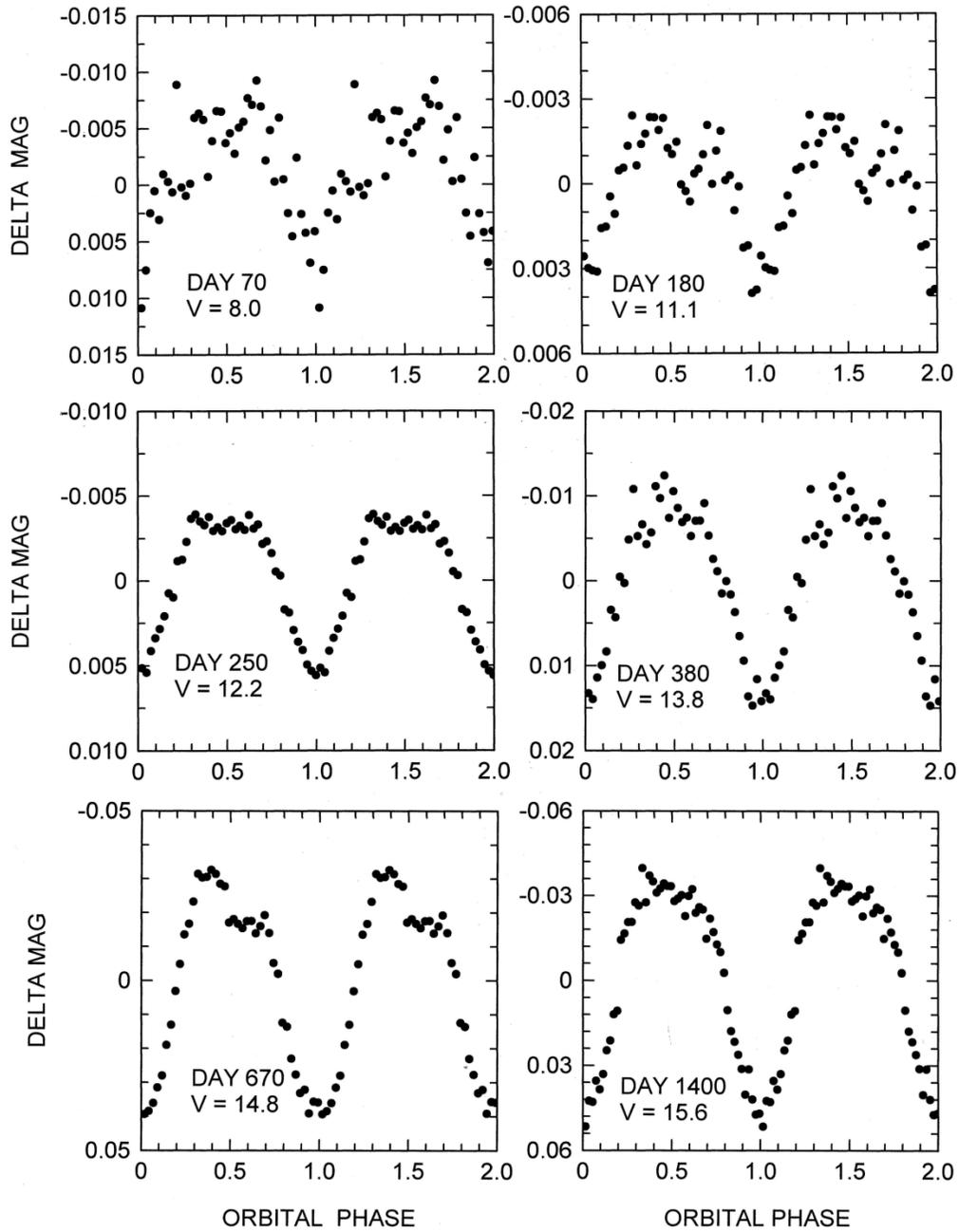

Figure 4. Orbital light curves at various points in the decline. Each is based on at least 20 orbits, and day numbers refer to the mid-point of the time series. The phase marker we have used — minimum light — appears to be stable within ~0.03 cycles. Despite appearances, the detection at day 70 is uncertain, since the peak in the power spectrum was too weak to be significant (this data was obtained near the 2011 solar conjunction).

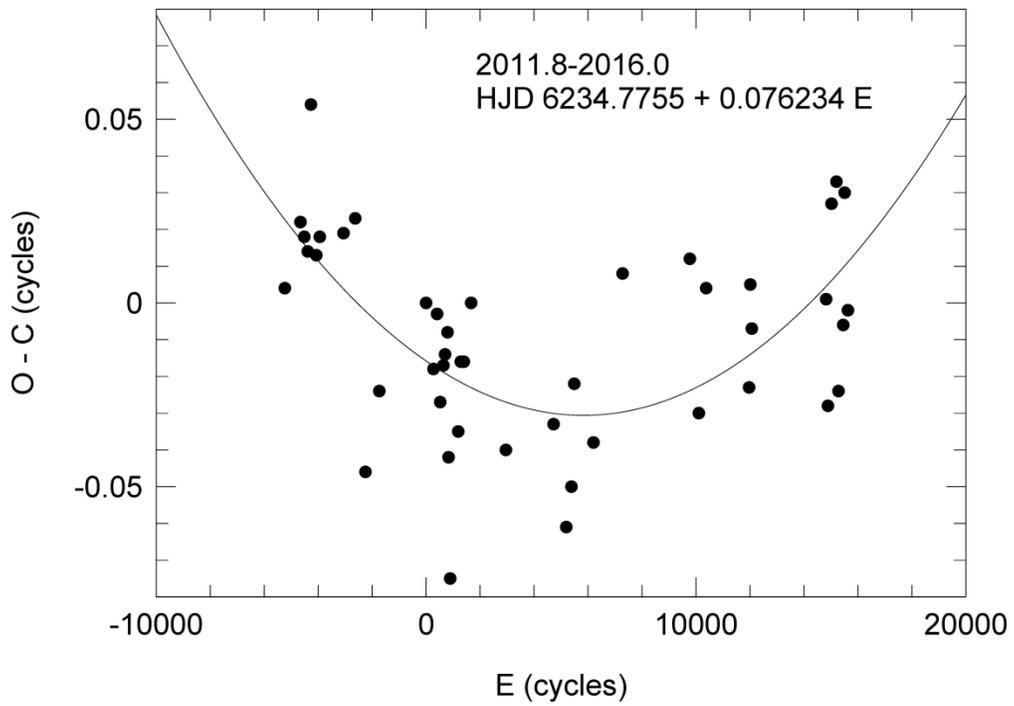

Figure 5.  O–C diagram for the timings after the orbital variation reappears.  The fitted curve indicates an increasing period, with $P/\dot{P} = 2.4(4)\times10^5$ years.

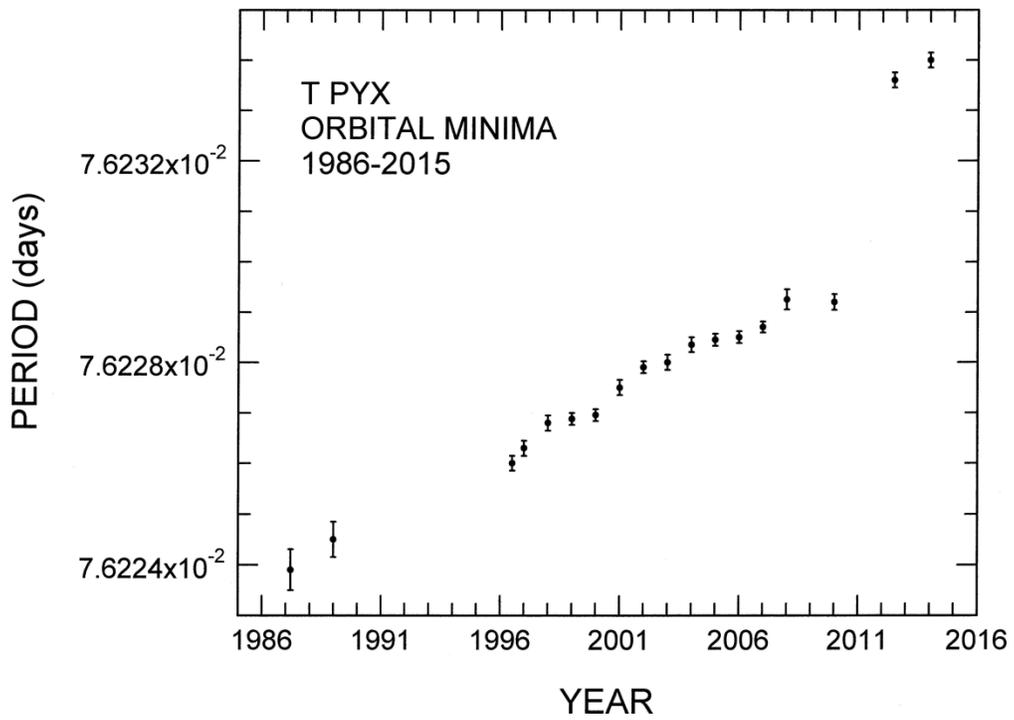

Figure 6. The variation of $P_{orb}$ during 1986–2015. Each point represents a two-year running mean. Day zero of the eruption occurred at 2011.4.

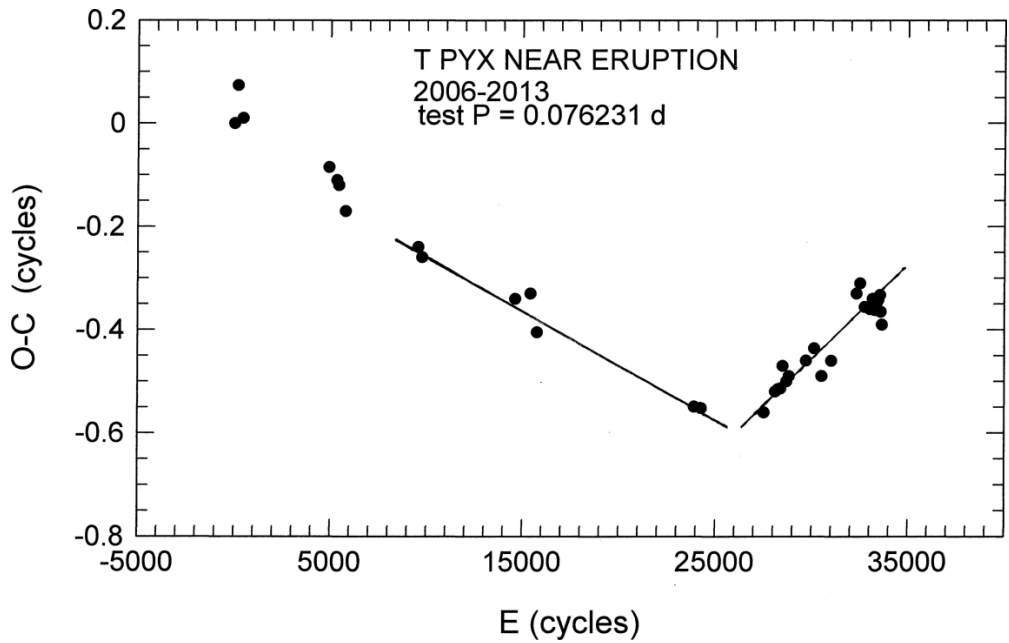

Figure 7. O–C diagram of orbital timings, relative to the test ephemeris in the figure, for several years before and after eruption. The straight lines define an apparent "V" vertex occurring at day 120±90; the change in slope indicates a change in period. The main point here is that the signal showed no discontinuity in phase — likely indicating that the periodic signal has the same origin, anchored in binary phase, before, after, and even *during* the eruption.

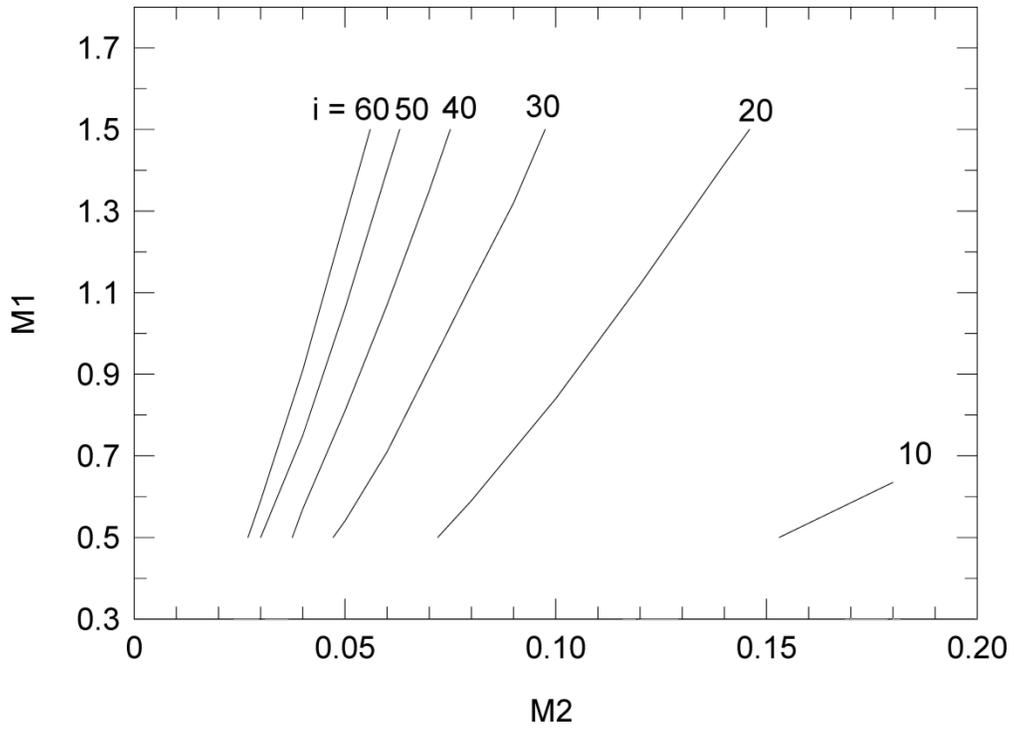

Figure 8. Constraint on the masses, for various choices of binary inclination $i$. The UKS measurement of $K_1 = v_1 \sin i = 18$ km/s is used. For the $i > 50°$ inclination (slightly) favored in this paper, $M_2$ must be less than 0.06 $M_\odot$.

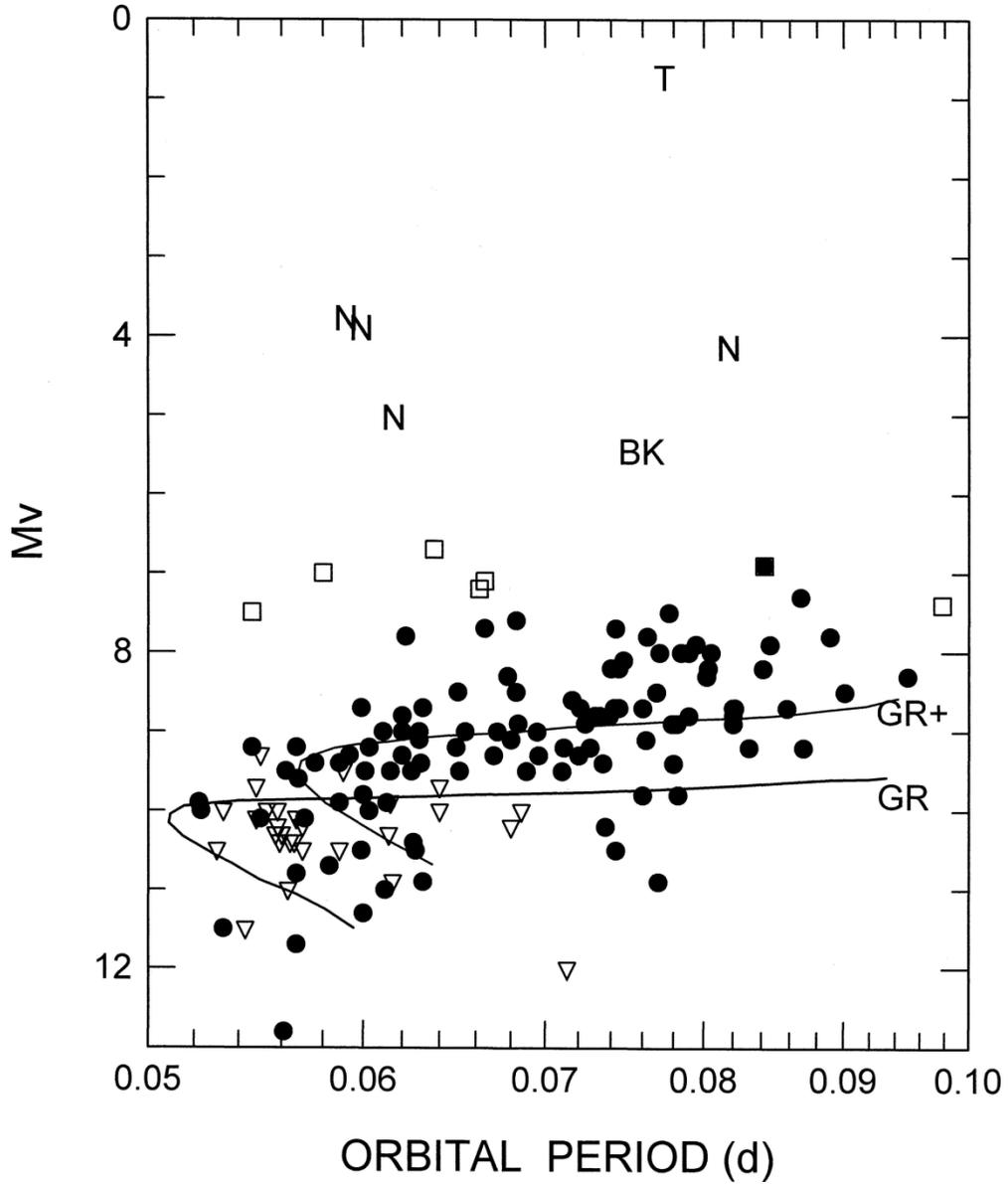

Figure 9. Time-averaged $M_V$ versus $P_{orb}$ for short-period CVs. The average error in <Mv>, usually dominated by distance uncertainty, is probably near 0.8 mag. This excludes actual nova eruptions, and therefore should predominantly represent *accretion* light. Dots are dwarf novae, triangles are upper limits for dwarf novae (usually because the recurrence time is not known), and the bold boomerang-shaped curve labelled GR is a theoretical main-sequence for CVs. (GR+ indicates an "enhanced GR", discussed in the text.) Stars labelled N are 20th century novae; "T" is T Pyx, and "BK" is a likely 2nd-century nova. The squares are ER UMa stars — dwarf novae which we interpret as millennia-old classical novae. In our interpretation of the aftermath of classical-nova eruptions, stars drop vertically down from $M_V = -7$, but with ever-increasing slowness, such that $dm/d(\log t) \approx 1$.